\documentclass{aa}
\usepackage[varg]{txfonts}
\pdfoutput=1

\usepackage{graphicx}

\title{Constraining the distribution of dark matter at the Galactic centre using the high-resolution Event Horizon Telescope}

\author{Thomas Lacroix
\and Joseph Silk}
\institute{UMR7095, Institut d'Astrophysique de Paris, 98 bis boulevard Arago, 75014 Paris, France\\
\email{[lacroix;silk]@iap.fr}}
\date{Received 16 November 2012 /
       Accepted 18 March 2013}

\begin{document}

\titlerunning{Constraining dark matter at the Galactic centre using the high-resolution Event Horizon Telescope}
\authorrunning{T. Lacroix \& J. Silk}

\abstract{We investigate constraints on the distribution of dark matter in the neighbourhood of the Galactic centre that may eventually be attained with the high-resolution Event Horizon Telescope (EHT). The shadow of a black hole in vacuum is used to generate a toy model describing how dark matter affects the size of the shadow of the supermassive black hole located at the Galactic centre. Observations by the EHT may constrain the properties of the dark matter distribution in a possible density spike around the black hole. Current uncertainties due to both the resolution of the telescope and the analysis of stellar orbits prevent one from discerning the effect of dark matter on the measured size of the shadow. The change in the size of the shadow induced by dark matter can be seen as an additional uncertainty in any test of general relativity that relies on using the angular size of the shadow to estimate the Schwarzschild radius of the black hole.}

\keywords{Dark matter - galaxy: centre - black hole physics}

\maketitle

\section{Introduction}

In the framework of the standard model of cosmology, dark matter comprises about $ 23 $ percent of the energy content of the universe. Consequently, unravelling the mystery of the nature of dark matter is one of the most challenging problems of modern cosmology. In this context, astronomical observations can provide significant constraints on the dark matter distribution. The dawning of a new era of telescopes with high resolution and sensitivity has widened the scope of precision experiments that may be used to set constraints on the properties of dark matter. In this work, we focus on one such experiment. We consider an experiment focusing on the Galactic centre, since the centre of the Milky Way has become a favoured region for searches for dark  matter signatures. The dark matter density is indeed expected to be higher in the neighbourhood of the Galactic centre, according to N-body simulations. Moreover, if dark matter is indeed made of weakly interacting particles, there is likely to be an enhancement of dark matter around the supermassive black hole Sgr A* located at the centre of the Milky Way.

The very high and still increasing resolution of the Event Horizon Telescope (EHT) is about to allow one to get a picture of the shadow of this black hole. Such an imaging survey may then be used to constrain the properties of a dark matter spike around the centre. This is of particular interest because such a geometrical measurement avoids any knowledge of the precise nature of the dark matter, such as its decay or annihilation properties. One needs a model describing how the dark matter around the centre modifies the size of the shadow to be able to make quantitative comparisons with future EHT pictures.

In Sect.~\ref{shadow of BH in vacuum}, we recall the notion of shadow of a black hole in a vacuum, before presenting a toy model in Sect.~\ref{shadow of BH + DM} that describes how the presence of dark matter around a black hole would affect the bending of light induced by the compact object and how this could be used to extract constraints on the properties of the dark matter distribution. Finally, we analyse the uncertainties involved in the process of inferring the properties of the black hole from the measurement of the size of the shadow.
 
\vspace{-0.2cm}
 
\section{Shadow of a Schwarzschild black hole} 
\label{shadow of BH in vacuum}

\begin{figure}[tbp]
\centering
\resizebox{8cm}{!}{\includegraphics{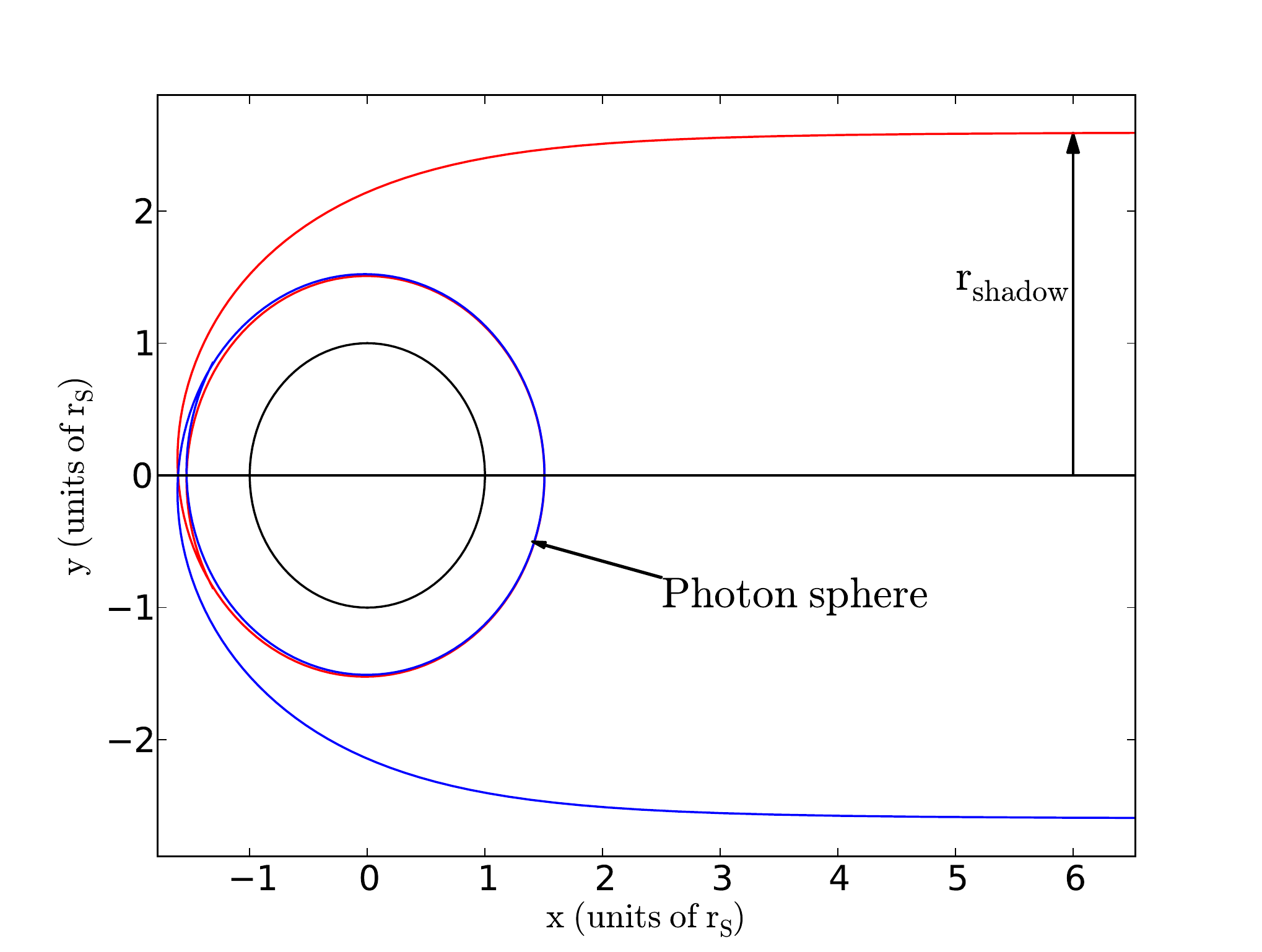}}
\caption{Shadow of a black hole. The radius of the shadow is the minimum impact parameter of a light ray escaping the black hole, so the shadow is a disk representing the black hole as seen by the observer. The circular orbit lies on the so-called photon sphere. The black circle represents the horizon.}
\label{shadow}
\end{figure}

The shadow of a black hole is closely related to the way that light rays are bent by the black hole and, more particularly, to the deflection angle. Starting from the Schwarzschild metric, which describes the geometry of a non-rotating black hole in a vacuum, one can show the relation between the impact parameter $b$ and the closest approach distance $ r_{\mathrm{m}} $ of a photon \citep{Weinberg}:
\begin{equation}
b = \dfrac{r_{\mathrm{m}}}{\sqrt{1 - \dfrac{r_{\mathrm{S}}}{r_{\mathrm{m}}}}}
\end{equation}
where $ r_{\mathrm{S}} $ is the Schwarzschild radius of the black hole. It turns out that the deflection angle features a divergence for a particular value of the impact parameter. This value defines the shadow of the black hole. More precisely, this divergence corresponds to light rays infinitely bent by the black hole and thus performing an infinite number of loops around it. Such orbits are unstable under small perturbations, which results in the photons eventually crossing the horizon and falling onto the singularity. Therefore, the shadow represents the minimum impact parameter of a photon escaping the attraction of the black hole. Conversely, it is the minimum impact parameter of a photon coming from infinity for it to end up on the photon sphere \citep{BHlensing}, as seen in Fig.~\ref{shadow}, so the shadow turns out to be the effective black disk seen by an observer looking at the black hole. The shadow border can thus be determined analytically since it corresponds to the minimum of the function $ b(r_{\mathrm{m}}) $. The photon sphere has a radius given by $ 1.5\ r_{\mathrm{S}} $. This value is obtained by minimizing $b$, and the corresponding value of the impact parameter yields the radius of the shadow: $ r_{\mathrm{shadow}} = 3\sqrt{3}/2\ r_{\mathrm{S}} \approx 2.6 \ r_{\mathrm{S}}$. 

Consequently, the shadow is the main observable feature of the black hole in a direct imaging survey, and this is what the EHT collaboration aims to observe in the near future, using the technique of Very Long Baseline Interferometry (VLBI). More precisely, the EHT project consists in phasing up millimetre and sub-millimetre telescopes scattered over the world, which together will form an effective high-resolution Earth-sized telescope \citep{2010evn..confE..53D}. The EHT array currently comprises the Submillimeter Telescope (SMT) in Arizona, the Combined Array for Research in Millimeter-wave Astronomy (CARMA) in California, and three telescopes in Hawaii: the Caltech Submillimeter Observatory (CSO), the James Clerk Maxwell Telescope (JCMT), and the Submillimeter Array (SMA) \footnote{see http://www.eventhorizontelescope.org/}. The network is soon to be complemented with the Atacama Large Millimeter/submillimeter Array (ALMA) in Chile. The data recorded simultaneously by all these telescopes are then processed by a dedicated supercomputer. So far, two baselines have been put together, namely the CARMA--SMT and the Hawaii--SMT baselines, in order to achieve an angular resolution of $ 58 $ microarcseconds ($ \rm \mu as $) at $ 230\ \rm GHz $. However, the angular Schwarzschild radius of the supermassive black hole Sgr A* is $ 10.3 \pm 1.24 \ \rm \mu as $ and thus the angular diameter of the shadow is of the order of $ 53.6 \ \rm \mu as $. This was obtained via $ r_{\mathrm{S}} = 2GM/c^{2} $, where $ c $ is the speed of light and $ G $ the gravitation constant, using the values of the distance of Sgr A* from the Sun, $ R_{\odot} = 8.28 \pm 0.33\ \rm kpc $, and the mass of Sgr A*, $ M = 4.3 \pm 0.36 \times 10^{6}\ M_{\odot}$ as given in \cite{BHmass}. The current resolution is therefore not yet sufficient to get an image of the shadow of the black hole, but it is quite close. Adding more baselines to the EHT array will thus improve the resolution so that the shadow of Sgr A* should be observed in the very near future. Moreover, the resolution is better at lower wavelengths, so future observations are going to be carried out at an additional, higher frequency, namely $ 345\ \rm GHz $. Consequently, the addition of ALMA, along with another baseline between the Plateau de Bure interferometer in France and the South Pole Telescope (SPT), will allow the EHT array to achieve a resolution of $ 15 \ \rm \mu as $ at $ 345\ \rm GHz $.

Imaging the shadow of a black hole will allow one to test the prediction of general relativity for the radius of the shadow and study accretion flows in the vicinity of the black hole. But it will also allow setting constraints on the properties of a hypothetical distribution of dark matter surrounding the supermassive black hole at the Galactic centre, and this is the object of the next section.

\vspace{-0.2cm}

\section{Shadow of a black hole surrounded by dark matter}
\label{shadow of BH + DM}

In this section we present a toy model describing how the shadow of a black hole is modified if the singularity is embedded in dark matter.

\vspace{-0.2cm}

\subsection{The modified metric}

To determine the behaviour of light rays, one must first derive the metric describing the geometry of spacetime induced by the black hole embedded in matter. The matter distribution is assumed to be spherically symmetric for the purpose of simplification. We start from the general form of a static and spherically symmetric metric \citep{Weinberg},
\begin{equation}
ds^{2} = B(r) dt^{2} - A(r) dr^{2} - r^{2} \left( d\theta^{2} + \sin^{2}\theta \ d\varphi^{2} \right), 
\end{equation}
and determine $ A $ and $ B $ using the Einstein equations. The dark matter halo is assumed to be collisionless and at rest. Under these assumptions, the first metric coefficient is given by
\begin{equation}
A(r) = \left( 1 - \dfrac{2 m(r)}{r} \right) ^{-1}
\end{equation}
where $ m(r) = \displaystyle \int_{0}^{r} \! 4 \pi r'^{2} \rho (r') \, \mathrm{d}r' $ is the enclosed mass in the sphere of radius $r$. The differential equation that rules the other unknown coefficient is derived by injecting the expression for $ A $ back into the Einstein equations and it reads as
\begin{equation}
\dfrac{B'}{B} = \dfrac{2 m(r)}{r^{2}} \left( 1 - \dfrac{2 m(r)}{r} \right) ^{-1}.
\end{equation}

\vspace{-0.4cm}

\subsection{Shadow of a black hole surrounded by a dark matter spike}

A dark matter density profile must now be chosen. Here we consider dark matter distributed according to a density spike around the central black hole. More precisely, the galaxy is believed to be embedded in a halo with a moderately steep profile ($ \rho \propto r^{-\beta} $ where $ 0 < \beta < 2 $) from N-body simulations (see \cite{Bringmann} or \cite{multi-wavelength} for reviews of dark matter at the Galactic centre). However, some models also predict a dark matter spike at the Galactic centre, that is a steep density profile within the central parsec \citep{Gondolo}. The slow growth of the black hole at the Galactic centre, within the halo, may have given rise to such a steep density profile for the dark matter component, owing to accretion of dark matter by the black hole: $ \rho \propto r^{-\gamma} $ where $ 2.25 \leqslant \gamma \leqslant 2.5 $. If the dark matter halo has a Navarro-Frenk-White (NFW) density profile ($ \rho \propto r^{-1} $ for small radii), then the spike profile turns out to be $ \rho \propto r^{-7/3} $ \citep{Gondolo}. 

In the remainder of this section, we neglect the contribution of the halo itself and focus on the gravitational lensing effects of the spike, which should be the dominant component of the system, as far as the shadow is concerned. Therefore, the spike is assumed to be surrounded by a vacuum in the calculations involving the shadow. A halo without a spike would have a negligible effect on the shadow, thanks to a less severe enhancement of the dark matter density at the centre.

We next consider the following density profile for the system comprising the black hole of mass $ M_{\mathrm{BH}} $ (equivalent to a point mass) and the dark matter spike of mass $ M_{\mathrm{spike}} $ and radius $ R $,
\begin{equation}
\rho (r) = M_{\mathrm{BH}} \dfrac{\delta (r)}{4 \pi r^{2}} + \alpha r^{-\gamma}
\end{equation}
where $ \gamma $ is taken as $ 7/3 $, and $ \alpha $ is determined by the normalization condition that the integral of the spike density be equal to the total mass of the spike. The uncertainty on $ \gamma $ leads to changes in the size of the shadow, which are negligible with respect to all other uncertainties by several orders of magnitude. Therefore with this model the differential equation for $B$ takes the form
\begin{equation}
\label{B'}
\dfrac{B'}{B} = \dfrac{\dfrac{2 M_{\mathrm{BH}}}{r^{2}} \left( 1 + q \left( \dfrac{r}{R} \right) ^{3 - \gamma} \right) }{1 - \dfrac{2 M_{\mathrm{BH}}}{r} \left( 1 + q \left( \dfrac{r}{R} \right) ^{3 - \gamma} \right)}
\end{equation}
where the mass ratio is $ q = M_{\mathrm{spike}}/M_{\mathrm{BH}} $. Considering that $R$ should be much larger than the Schwarzschild radius $ r_{\mathrm{S}} = 2 M_{\mathrm{BH}} $, and $q$ should be smaller than 1, the second term in each bracket is going to be a perturbative term for small radii ($ r \sim r_{\mathrm{S}} $), relevant to deriving the radius of the shadow.
Consequently, equation (\ref{B'}) can be approximately integrated using the fact that the numerator on the right-hand side is the derivative of the denominator, except for the small term containing $q$ and $R$. This leads to
\begin{equation}
B \approx B_{0} \left( 1 - \dfrac{r_{\mathrm{S}}}{r} \left( 1 + q \left( \dfrac{r}{R} \right) ^{3 - \gamma} \right) \right) 
\end{equation}
for $ r < R $, where $ B_{0} $ is a constant that can be obtained using the expression of the metric coefficient $B$ in a vacuum: $ B(r) = 1 - r_{\mathrm{S,tot}}/r $ for $ r > R $, where $ r_{\mathrm{S,tot}} = r_{\mathrm{S}}(1+q) $ is the total Schwarzschild radius corresponding to $ M_{\mathrm{tot}} = M_{\mathrm{BH}} + M_{\mathrm{spike}} $. Therefore continuity of the metric at the boundary of the spike $ r = R $ leads to $ B_{0} = 1 $. In the following, the metric used to describe the geometry induced by a black hole surrounded by a dark matter spike will be determined by the coefficients
\begin{align}
\label{A&B}
A(r) &= \left( 1 - \dfrac{r_{\mathrm{S}}}{r} \left( 1 + q \left( \dfrac{r}{R} \right) ^{3 - \gamma} \right) \right) ^{-1} \nonumber \\
B(r) &\approx A(r)^{-1}
\end{align}
where the first equation is exact, and the second equation holds for parameters of the spike such that the perturbative approach is valid. More precisely, to make sure that this approximation was reasonable, we numerically solved equation (\ref{B'}) using a Runge-Kutta scheme to compute the actual function $ B(r) $, given the boundary condition $ B(R) = 1 - r_{\mathrm{S,tot}}/R $. Then we computed the product of functions $ A(r) $ given in (\ref{A&B}) and $ B(r) $ obtained numerically. It turned out to be equal to $ 1 $ with an error of $ 10^{-3} $ for $ q = 0.1 $ and $ R = 10^{3}\ r_{\mathrm{S}} $, which are sensible limits from a physical point of view (since the spike is predicted to be both large and light with respect to the black hole). Such an error will be insignificant because the radius of the shadow will be given with a much greater uncertainty than one percent (see below). Moreover, these values of $ q $ and $ R $ are the very limits of this perturbative approach, and for most values of the parameters, the error is actually even smaller than $ 10^{-3} $.

With these approximations, the metric can be derived analytically for the density profiles considered here. Therefore, the approximations detailed above allow us to pursue the calculation and obtain an analytic expression for the size of the shadow $ r_{\mathrm{shadow}} $. This allows for faster numerical calculations. Then, constraints on the properties of the system can be extracted from EHT pictures by measuring the deviation of the actual shadow with respect to the expected shadow of a black hole in vacuum. Consequently, uncertainties on the measurement of $ r_{\mathrm{shadow}} $ are going to play a crucial part.

First of all, one needs to compute the size of the shadow for different values of $q$ and $R$. Similar to the case of a black hole in vacuum, the impact parameter $b$ as a function of the closest approach distance $ r_{\mathrm{m}} $ is given by
\begin{equation}
b(r_{\mathrm{m}}) = \dfrac{r_{\mathrm{m}}}{\sqrt{B(r_{\mathrm{m}})}}.
\end{equation}
Then the radius of the shadow is given by minimizing this function with respect to $ r_{\mathrm{m}} $. The impact parameter assumes its minimum for a value of $ r_{\mathrm{m}} $ such that
\begin{equation}
\label{r_m_min}
r_{\mathrm{m}}^{\mathrm{(min)}} = \dfrac{3}{2} r_{\mathrm{S}} \left(1 + q \left( \dfrac{r_{\mathrm{m}}^{\mathrm{(min)}}}{R} \right) ^{3 - \gamma} \right).
\end{equation}
This equation can be solved perturbatively, considering that $ q \left( r_{\mathrm{m}}^{\mathrm{(min)}}/R \right) ^{3 - \gamma} \ll 1 $ for realistic models. We solve it by iteration, starting from the zeroth order and injecting the solution back into the right-hand side of equation (\ref{r_m_min}). Only four iterations are required to retrieve the solution with a precision higher than $ 10^{-3} $. This gives an analytic expression for the radius of the shadow $ r_{\mathrm{shadow}} = b(r_{\mathrm{m}}^{\mathrm{(min)}}(q,R)) $ as a function of $q$ and $R$. 

\begin{figure}[tbp]
\centering
\resizebox{8cm}{!}{\includegraphics{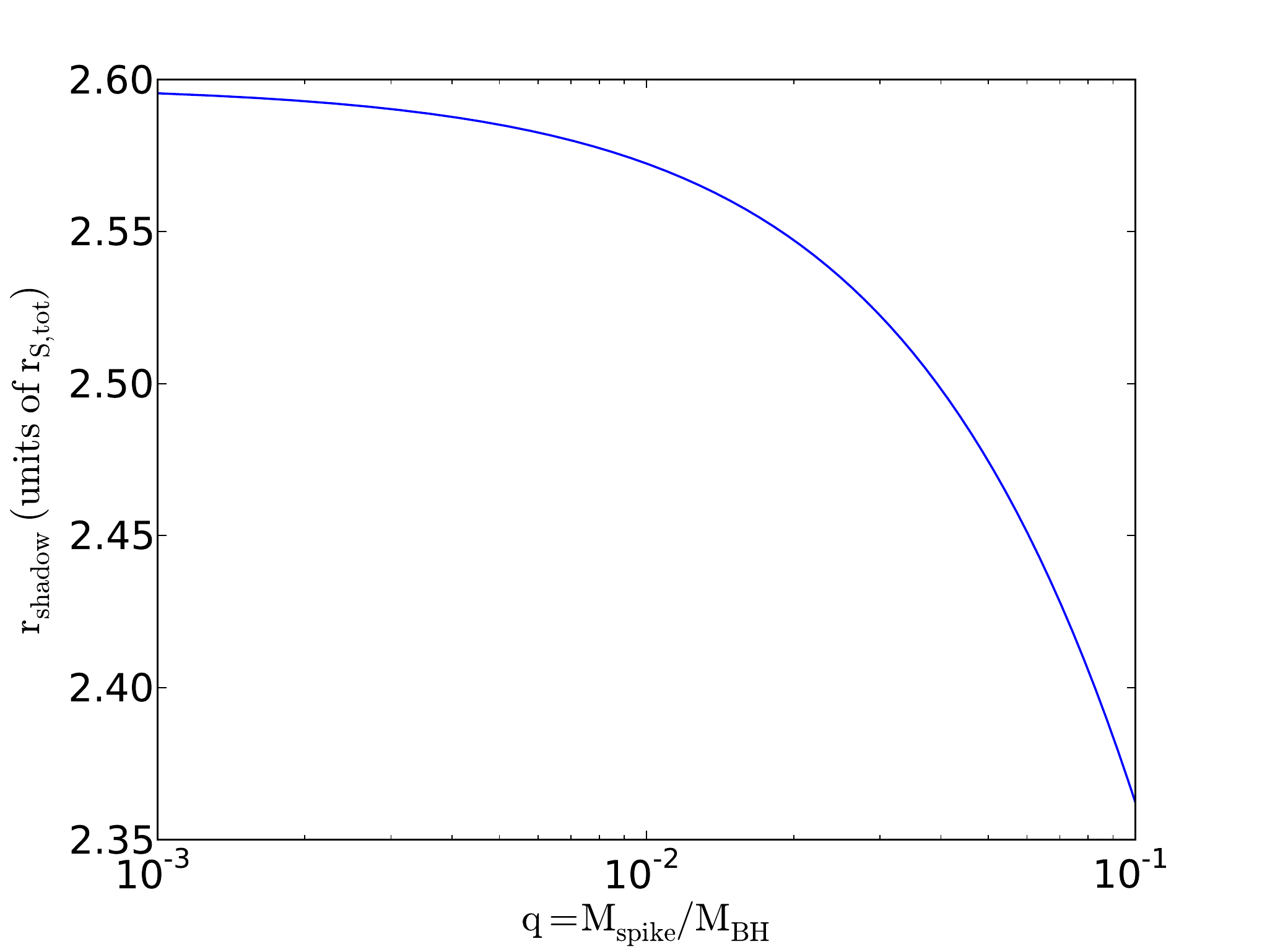}} 
\caption{Radius of the shadow $ r_{\mathrm{shadow}} $ as a function of the mass ratio $ q = M_{\mathrm{spike}}/M_{\mathrm{BH}} $. $ r_{\mathrm{shadow}} $ is expressed in units of the total Schwarzschild radius $ r_{\mathrm{S,tot}} $.}
\label{r_shadow_vs_q}
\end{figure}

It turns out that there is a degeneracy in $ R $. More precisely, the size of the shadow does not depend on the radius of the spike for $ R $ larger than $ 10^{3} \ r_{\mathrm{S}} $. Here we do not consider any lower values of the radius of the spike because $ R $ is often predicted to be large with respect to $ r_{\mathrm{S}} $ (for instance $ R \sim 1\ \rm pc $ according to \citet{Merritt} and $ r_{\mathrm{S}} \sim 10^{-7}\ \rm pc $).

\vspace{-0.2cm}

\subsection{Resulting constraints and uncertainties}

Considering the degeneracy in $ R $, the best approach is to plot the resulting curve of $ r_{\mathrm{shadow}} $ as a function of the mass ratio $ q $, as shown in Fig.~\ref{r_shadow_vs_q}. It is important to note that here $ r_{\mathrm{shadow}} $ is expressed in units of the total Schwarzschild radius $ r_{\mathrm{S,tot}} $ corresponding to the total mass $ M_{\mathrm{tot}} = M_{\mathrm{BH}} + M_{\mathrm{spike}} = M_{\mathrm{BH}} (1 + q)$. This is the only quantity known from previous studies of the motion of stars at the Galactic centre: $ M_{\mathrm{tot}} = 4.3 \pm 0.36 \times 10^{6}\ M_{\odot}$ \citep{BHmass}. This value is in general taken as the black hole mass, but dynamical studies cannot separately measure $ M_{\mathrm{spike}} $ and $ M_{\mathrm{BH}} $, so we consider that the mass given in the  literature is the total mass in the framework of a black hole surrounded by a dark matter spike. Therefore, this allows one to switch from the unknowns $ R $, $ M_{\mathrm{BH}} $, and $ M_{\mathrm{spike}} $ to $ R $, $ M_{\mathrm{tot}} $, and $ q $, where $ M_{\mathrm{tot}} $ is already known.

\begin{figure}[tbp]
\centering
\resizebox{8cm}{!}{\includegraphics{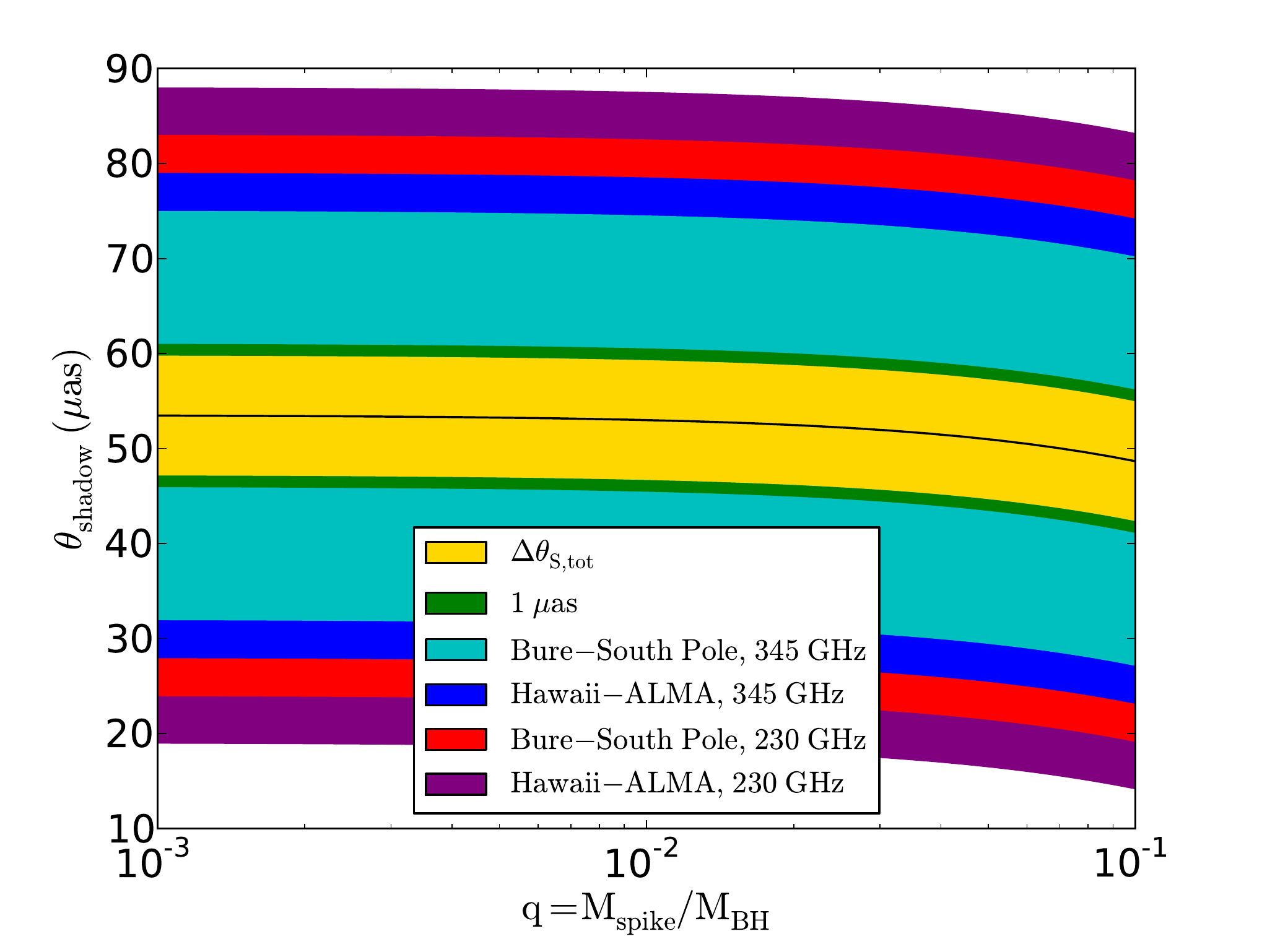}} 
\caption{Angular diameter of the shadow as a function of the mass ratio $ q = M_{\mathrm{spike}}/M_{\mathrm{BH}} $ (black solid line). The yellow shaded area is the filled contour corresponding to the uncertainty on the shadow due to the error on the total angular Schwarzschild radius, $ \Delta \theta _{\mathrm{S,tot}} = 6.43\ \rm \mu as $. The other contours show the uncertainty due to resolution, for the baselines (Hawaii--ALMA and Bure--SPT), which allow the array to achieve its best resolution, and for two observing frequencies. The $ 1\ \rm \mu as $ resolution is an optimistic case.}
\label{theta_shadow_vs_q}
\end{figure}

Considering that the size of the shadow obtained by solving (\ref{r_m_min}) is expressed in units of the Schwarzschild radius $ r_{\mathrm{S}} $ of the black hole alone, and as $ r_{\mathrm{S,tot}} = r_{\mathrm{S}} (1 + q) $, the value of $ r_{\mathrm{shadow}} $ given by the calculation had to be divided by $ (1+q) $, in order to be expressed in terms of $ r_{\mathrm{S,tot}} $ (which is a known length scale, unlike $ r_{\mathrm{S}} $). Moreover, expressing $ r_{\mathrm{shadow}} $ in terms of $ r_{\mathrm{S,tot}} $ allows us to break a degeneracy, since $ r_{\mathrm{shadow}} $ in units of $ r_{S} $ is very close to $ 2.6 $ for all values of $ q $ and $ R $ in the region of interest. The aim is then to set constraints on the remaining unknowns $ q $ and $ R $. 

The quantity of interest from the observational point of view turns out to be the angular diameter of the shadow $ \theta _{\mathrm{shadow}} = 2 r_{\mathrm{shadow}} / R_{\odot} $. Shown in Fig.~\ref{theta_shadow_vs_q} is the angular diameter of the shadow derived from the previous calculation. This graph provides a clear way of constraining the mass ratio from the value of the radius of the shadow. In this study $ q $ ranges from $ 10^{-3} $ to $ 10^{-1} $. Going down to lower values would be useless, since $ r_{\mathrm{shadow}}(q) $ is almost constant below $ 10^{-3} $. Moreover, the spike is likely to have a perturbative effect, so $ q $ should be small enough. That is why $ q $ does not go above $ 10^{-1} $ here. Finally, the range considered here shows the most compelling variations of the size of the shadow: $ \theta _{\mathrm{shadow}} $ goes down by $ 5\ \rm \mu as $ between $ 10^{-3} $ and $ 10^{-1} $, which sets the minimum uncertainty required to discern an effect of dark matter. The size of the shadow is in fact mostly sensitive to values of the mass ratio $ q $ greater than $ 10^{-2} $. Consequently, measuring the size of the shadow in principle allows one to distinguish between high values of the mass ratio. However, the uncertainty on the shadow due to the error on the total angular Schwarzschild radius, $ \Delta \theta _{\mathrm{S,tot}} = 6.43\ \rm \mu as $, must be taken into account. The angular size of the shadow in this model is indeed derived in terms of the total angular Schwarschild radius $ \theta _{\mathrm{S,tot}} = r _{\mathrm{S,tot}} / R_{\odot} $. This uncertainty comes from the error on the measurement of the distance of Sgr A* from the Sun, $ R_{\odot} = 8.28 \pm 0.33\ \rm kpc $, and the total mass, $ M_{\mathrm{tot}} = 4.3 \pm 0.36 \times 10^{6}\ M_{\odot}$ \citep{BHmass}. Consequently, it turns out that, even before taking the resolution of the EHT into account, the uncertainty due to the mass and distance of Sgr A* is already greater than the width of the range of values of $ \theta _{\mathrm{shadow}} $. To observe an effect of dark matter, the sum of the uncertainty on $ \theta _{\mathrm{S,tot}} $ and the error due to resolution should be smaller than $ 5\ \rm \mu as $.

Shown in Fig.~\ref{theta_shadow_vs_q} are only the best resolutions achievable by adding the Hawaii--ALMA and Plateau de Bure--SPT baselines to the current baselines (CARMA--SMT and Hawaii--SMT), for two observing frequencies at $ 230\ \rm GHz $ and $ 345\ \rm GHz $. Finally, we consider a resolution of $ 1\ \rm \mu as $, which would be ideal to carry out a precise measurement of the size of the shadow and observe a possible change induced by dark matter. For now, a resolution as high as $ 15\ \rm \mu as $, despite being huge by astronomical standards, will only allow one  to see the shape of the shadow. This is the primary goal of the EHT since the collaboration aims to shed light on physical processes taking place in the vicinity of the black hole, such as accretion flows. The following step is to measure the size of the shadow to obtain a direct estimate of the angular Schwarzschild radius of the black hole and possibly see the effect of dark matter. However, with a resolution of $ 15\ \rm \mu as $, the size of the shadow will be measured with an error of almost 30 percent. Therefore, with such a large error, discerning a deviation from the value of the size of the shadow in vacuum is going to prove a challenging task, but even measuring the size of the shadow with high enough precision is going to be difficult. As a matter of fact, the EHT is going to be able to carry out a direct measurement of the total Schwarzschild radius, but with an uncertainty of $ 15\ \rm \mu as $, i.e.\ larger than the error coming from the measurement using stellar orbits. The existing constraints on $ \theta _{\mathrm{S,tot}} $ are thus not going to be improved from the observation of the shadow with a resolution of $ 15\ \rm \mu as $. Nevertheless, this resolution planned for the near future is certainly not the ultimate goal, and it can surely be improved even more by adding still longer baselines, with potential EHT sites in New Zealand and Africa.

\begin{figure}[tbp]
\centering
\resizebox{8cm}{!}{\includegraphics{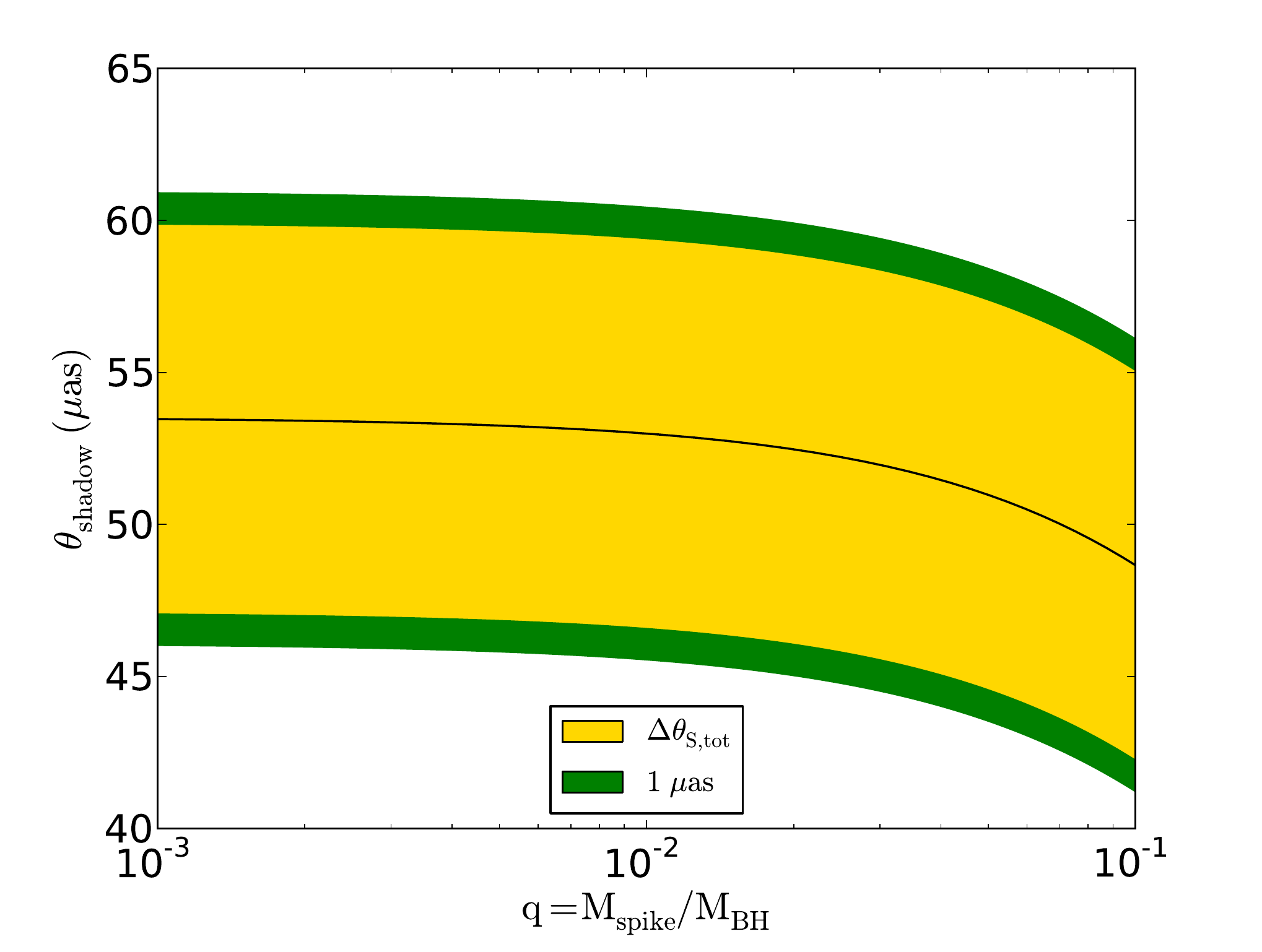}}
\caption{Angular diameter of the shadow as a function of the mass ratio $ q = M_{\mathrm{spike}}/M_{\mathrm{BH}} $ (black solid line) with only uncertainties coming from the error on the total angular Schwarzschild radius (yellow) and an ideal resolution of $ 1\ \rm \mu as $ (green).}
\label{theta_shadow_vs_q_smallest_uncertainties}
\end{figure}

We now focus on the constraints on the dark matter spike. On the one hand, the angular size of the shadow can only be known with an error over $ 7\ \rm \mu as $, due to the uncertainty on $ \theta _{\mathrm{S,tot}} $, and even for a resolution as high as $ 1\ \rm \mu as $ (Fig.~\ref{theta_shadow_vs_q_smallest_uncertainties}). As a result, the error on the size of the shadow is actually too large for $ q $ to be constrained. The uncertainty on $ \theta _{\mathrm{S,tot}} $ turns out to be critical, and even increasing the resolution of the EHT much more would not allow us to set stringent constraints on $ q $. Therefore, placing constraints on the dark matter distribution not only requires improved resolution, but also smaller uncertainties on the total mass and the distance of Sgr A*.  

On the other hand, the degeneracy in $ R $ means that one cannot set constraints on the size of the spike directly from this study of the shadow. In fact, there is a relation between $ q $ and $ R $, but it is independent of the calculation of the shadow. So far, for the purpose of determining how dark matter affects the shadow, we have neglected the smooth distribution (with an NFW profile) surrounding the spike, since its effect on the shadow is negligible. However, this underlying distribution can be used to derive a relation between the mass ratio and the radius of the spike, using the continuity of the density at the boundary of the spike: 
\begin{equation}
\label{relation_M_spike_R}
R = \left( \dfrac{q M_{\mathrm{tot}}}{6 \pi \rho_{0} R_{\mathrm{gal}} (1+q)}\right) ^{1/2}
\end{equation}
with $ \rho_{0} = \rho_{\odot} \dfrac{R_{\odot}}{R_{\mathrm{gal}}} \left( \dfrac{R_{\odot}}{R_{\mathrm{gal}}} + 1 \right) ^{2} $ where $ R_{\mathrm{gal}} = 20 \ \rm kpc $ is the radius characterising the dark matter halo, $ R_{\odot} = 8.28 \ \rm kpc $ is the distance of the solar system from the Galactic centre and $ \rho_{\odot} = 0.3 \ \rm GeV \ cm^{-3} $ is the local dark matter density at the Sun. Here we have used $ R \ll R_{\mathrm{gal}} $ to obtain a simplified expression. Since $ q $ cannot be properly constrained at the moment, the radius of the spike cannot be constrained either, but equation (\ref{relation_M_spike_R}) provides a way of translating future constraints on $ q $ into constraints on $ R $, with an uncertainty given by the uncertainties on $ q $, $ M_{\mathrm{tot}} $ and $ R_{\odot} $.

Although a direct measurement of the effect of dark matter is made more difficult by large uncertainties, quantifying the change in the shadow induced by a dark matter spike is useful as it turns out to be an additional source of uncertainty when testing the predictions of general relativity by measuring the size of the shadow.

\vspace{-0.2cm}

\subsection{Comments on some approximations made in this study}

First of all, throughout this work we have neglected the effect of stars located around the Galactic centre on the bending of light and, more precisely, on the size of the shadow of the black hole. This seems indeed a sensible approximation since stars around the Galactic centre form a mass distribution that extends much further than the dark matter spike. Therefore, even if the luminous mass around the Galactic centre is significant, its effect on gravitational lensing is actually negligible due to its being too widespread, as can be seen from the previous sections. Moreover, there is much less severe enhancement of the density of stars at the centre than what is expected for dark matter \citep{Merritt}. Stars behave like a background mass distribution, similarly to the dark matter halo in which the whole galaxy is embedded, and thus make a negligible contribution as far as changes in the size of the shadow are concerned. There is also a central nuclear star cluster, which will scatter dark matter particles and flatten the density profile to a slope of $r^{-3/2}$ \citep{Gnedin}. This is important outside a radius where the enclosed mass of stars is comparable to the black hole mass, but the dark matter profile in the region of interest for the shadow will not be affected.

Then we made approximations related to the properties of the black hole itself. A Schwarzschild black hole is indeed an idealized case, and from a more realistic point of view, the supermassive black hole at the Galactic centre should be rotating. The effect of the spin of such a Kerr black hole is to compress the shadow border on one side and to shift its position \citep{BHlensing}. Therefore, it is still possible to compute the radius of the shadow, which remains the same as for the Schwarzschild black hole for the non-compressed part. Consequently, it is not necessary in principle to consider the more complicated case of a rotating black hole to analyse the effect of a dark matter distribution on the size of the shadow. However, the orientation of the spin axis of the black hole is unknown, so this may create additional uncertainty in our model. Further work will be needed to clarify the influence of the spin on the shadow of a black hole surrounded by dark matter, but this was not the primary purpose of the present study.

Finally, it is worth noticing that this study of the shadow does not depend on any assumption about the nature of dark matter. The model only depends on a dark matter mass distribution, regardless of the nature of the dark matter. Therefore, if a positive effect on the shadow were to be detected, this may be a sign that dark matter is not an artefact of the way observations are interpreted.

\section{Conclusion}

In this work, we have focused on an original way of setting constraints on the properties of dark matter by observing the Galactic centre. The study of the shadow of the central black hole embedded in a spike of dark matter in principle allows one to constrain the properties of the spike, using future observations carried out by the Event Horizon Telescope. The EHT is indeed close to reaching a high enough resolution to take a picture of the shadow. We have presented a model that describes the effect of a dark matter spike on the shadow of the black hole Sgr A*. Values predicted by this model are then to be compared to the actual value of the size of the shadow measured by the EHT, in order to discern the effect of dark matter and constrain the mass and the radius of the spike. We quantified the various uncertainties involved in the problem. It turns out that the uncertainties on the mass and distance of Sgr A* that translate into an error $ \Delta \theta _{\mathrm{S,tot}} $ on the shadow, along with the resolution of the EHT array, make it impossible to set relevant constraints on the dark matter distribution. Consequently, increasing the resolution is not sufficient to take advantage of a direct observation of the shadow, but the focus should also be on reducing the uncertainties on $ \theta _{\mathrm{S,tot}} $. This is crucial when trying to discern deviations from the case of a black hole in vacuum. Nevertheless, this apparent drawback of the method can be turned into an advantage, because this method allows one to compute the scale of variation of the size of the shadow, $ 5\ \rm \mu as $. This proves to be an additional uncertainty when estimating the angular Schwarzschild radius of Sgr A* from the size of the shadow, which is one goal of the EHT.

\begin{acknowledgements}

We would like to thank C\'{e}line B\oe hm, Alexander Belikov, and Timur Delahaye for valuable comments. We are also indebted to Kaiki Inoue, Pierre Fleury, and Cl\'{e}ment Ranc for fruitful discussions. This research was supported in part by ERC project 267117 (DARK) hosted by Universit\'{e} Pierre et Marie Curie - Paris 6.

\end{acknowledgements}

\bibliographystyle{aa} 
\bibliography{shadow_paper_biblio}

\begin{thebibliography}{9}
\expandafter\ifx\csname natexlab\endcsname\relax\def\natexlab#1{#1}\fi

\bibitem[{Bozza(2010)}]{BHlensing}
Bozza, V. 2010, General Relativity and Gravitation, 42, 2269

\bibitem[{Bringmann \& Weniger(2012)}]{Bringmann}
Bringmann, T. \& Weniger, C. 2012, Physics of the Dark Universe, 1, 194

\bibitem[{Doeleman(2010)}]{2010evn..confE..53D}
Doeleman, S. 2010, in 10th European VLBI Network Symposium and EVN Users
  Meeting: VLBI and the New Generation of Radio Arrays (PoS)

\bibitem[{{Gillessen} {et~al.}(2009){Gillessen}, {Eisenhauer}, {Fritz},
  {et~al.}}]{BHmass}
{Gillessen}, S., {Eisenhauer}, F., {Fritz}, T.~K., {et~al.} 2009, Astrophys. J.
  Lett., 707, L114

\bibitem[{Gnedin \& Primack(2004)}]{Gnedin}
Gnedin, O. \& Primack, J. 2004, Phys. Rev. Lett., 93, 061302

\bibitem[{Gondolo \& Silk(1999)}]{Gondolo}
Gondolo, P. \& Silk, J. 1999, Phys. Rev. Lett., 83, 1719

\bibitem[{Merritt(2010)}]{Merritt}
Merritt, D. 2010, in Particle Dark Matter: Observations, Models and Searches
  (Cambridge University Press)

\bibitem[{Regis \& Ullio(2008)}]{multi-wavelength}
Regis, M. \& Ullio, P. 2008, Phys. Rev. D, 78, 043505

\bibitem[{Weinberg(1972)}]{Weinberg}
Weinberg, S. 1972, {G}ravitation and {C}osmology (Wiley, New York)

\end{thebibliography}

\end{document}